\documentclass[acmtog,nonacm]{acmart}

\usepackage{amsmath,booktabs}
\usepackage{algorithm}
\usepackage{algpseudocode}
\usepackage{amsthm}
\newtheorem{proposition}{Proposition}
\usepackage{placeins}
\usepackage{microtype}

\AtBeginDocument{%
  }

\setcopyright{none}
\copyrightyear{2026}
\acmYear{2026}
\acmDOI{}
\acmISBN{}

\title{Incremental Matrix Assembly for Preplanned Dynamic Tetrahedral Meshes}

\author{Manish Acharya}
\affiliation{%
  \institution{Vanderbilt University}
  \country{USA}}
\email{manish.acharya@vanderbilt.edu}

\author{David Hyde}
\affiliation{%
  \institution{Vanderbilt University}
  \country{USA}}
\email{david.hyde.1@vanderbilt.edu}

\renewcommand{\shortauthors}{Acharya and Hyde}

\begin{abstract}
Local topology edits require updating assembled operators even when most of a tetrahedral mesh remains unchanged.
We implement and evaluate incremental matrix assembly for preplanned edits, assuming a fixed pool of candidate elements with precomputable contributions.
A preallocated sparse layout allows operator values to be updated from edited elements without rebuilding the sparsity structure.
We compare full reassembly, local recomputation, and persistent incremental updates using a shared layout.
Across tested meshes with up to 460k tetrahedra, incremental assembly achieves 6--27\(\times\) assembly speedups over a vectorized full-reassembly baseline for a proxy graph Laplacian and 2--17\(\times\) for linear elasticity.
Frame times decrease by 1--3\% when the linear solve dominates and by 18--20\% in the tested synthetic dynamic elasticity loop with shorter solves.
Local Recompute often achieves comparable frame times.
For fixed time-step coefficients, directly maintaining the combined elasticity system also avoids a global matrix-combination pass.
The Laplacian matches full reconstruction exactly in every tested frame; elasticity differences remain at rounding scale, reaching a maximum entry difference of \(4.2\times10^{-16}\) over 5,000 frames.
\end{abstract}



\begin{document}

\maketitle

\section{Introduction}

Tetrahedral simulations in computer graphics can activate, deactivate, or subdivide elements over time \cite{obrien1999brittle,molino2004virtual,wicke2010dynamic}.
After such an edit, the assembled matrix must be updated, even when most of the mesh is unchanged.
Reassembling the whole matrix is simple, but its cost grows with the whole mesh even when the edit is small.

Some simulation pipelines permit a fixed pool of candidate elements.
An object may be pre-fractured along fixed fragment boundaries, so the simulation only decides which elements separate and when \citep{sellan2023breaking,muller2013real}.
Adaptive methods may generate a refinement hierarchy ahead of time, so that every coarse tetrahedron has a known set of possible children \citep{koschier2014adaptive,KOSSACZKY1994275,maubach1995local,arnold2000locally,mitchell201630}.
In surgical simulation, the elements that could be merged may be known even if the timing is not \cite{bielser2003state,molino2004virtual}.
We call a simulation \emph{preplanned} when all elements it could ever use can be allocated before it starts.

This paper asks a simple question: in a preplanned simulation, how much time can we save by updating the matrix after each edit instead of rebuilding it?

Our answer has two parts.
Assembly savings are substantial, and the cost of updating one operator follows the size of the edit.
Frame-time gains are smaller, because they are limited by the rest of the frame, especially the linear solve.
A separate refinement replay tests whether direct assembly savings persist when initialization and the cost of applying the reserved matrix are included.

Our approach combines two known ideas.
We store the matrix in one fixed sparse layout that covers every entry any candidate element could use, so the sparsity structure need not be rebuilt.
We also keep a small amount of state between frames, so that each edit can be applied without looking at unchanged elements.
For the proxy Laplacian this state is a count of the active tetrahedra on each edge.
For linear elasticity it is the running sum of element contributions at each matrix entry \cite{sifakis2012fem}.
Our experiments use scripted edits and synthetic loads to isolate assembly costs and measure their effect on the surrounding solve loop.

Our main contributions are:
\begin{itemize}
  \item an implementation of incremental assembly for removing, subdividing, and reactivating tetrahedra on a fixed pool of candidate elements, without rebuilding the sparsity structure during the simulation;
  \item a comparison against a full rebuild and a local recomputation method, all written with the same numerical tools, with matrix-parity checks run separately from timing;
  \item measurements of assembly time, frame time, setup cost, memory, accuracy, and long-run rounding error across five meshes of three objects \cite{cutler2004simplification}, with paired significance tests, plus a separate refinement replay that includes initialization and matrix application costs.
\end{itemize}

\section{Related Work}

\subsection{Dynamic Topology in Simulation}

Topology change has long been a central problem in physically based animation.
\citet{obrien1999brittle} drove fracture from stress analysis on volumetric meshes, and \citet{molino2004virtual} handled cutting and separation with virtual nodes.
\citet{wicke2010dynamic} integrated local remeshing into elastoplastic simulation.
\citet{koschier2014adaptive} introduced reversible tetrahedral refinement for brittle fracture, which is the kind of prepared hierarchy our method can use.
\citet{hahn2015high} used boundary elements for high-resolution brittle fracture, \citet{ferguson2023remeshing} remeshed within a single time step for contacting elastodynamics, and \citet{fan2022brittle} simulated brittle fracture with material points.
\citet{wu2015survey} survey the simulation of cuts in deformable bodies.
Our focus is matrix maintenance following supplied topology edits.

\subsection{Sparse Matrix Reuse and Assembly}

Sparse systems often benefit from local modification instead of repeated reconstruction.
\citet{davis1999modify} modify a sparse Cholesky factorization at a cost tied to the part that changes, and CHOLMOD brought these updates into widely used software \cite{chen2008cholmod}.
\citet{zarebavani2025parth} reuse Cholesky reordering when the sparsity pattern changes.
\citet{yeung2018amps} maintain FEM systems under real-time cutting with augmented matrices and Schur complement updates, which avoids refactorization.
\citet{lee2007progressive} update the inverse stiffness matrix after a tetrahedron is removed in surgical cutting, recomputing only the affected nodes.
These methods work on the factorization, the inverse, or the solved system.
We work one layer earlier, on the assembled matrix, so this approach suits iterative solvers and could in principle feed the methods above.

Updating only what changed is also an established idea in assembly.
\citet{deng2022reduced}, for example, update a small number of stiffness entries in each Newton iteration of a plasticity model instead of reconstructing the matrix.
Other work speeds up assembly itself, for example on GPUs \cite{cecka2011assembly,muellerroemer2018gpu}, and reserving matrix memory ahead of time is standard advice in sparse libraries such as PETSc \cite{balay2023petsc}.

We evaluate these established assembly and preallocation techniques on a pool of candidate tetrahedra, and we measure when they help and when they do not.

\section{Method}

The method takes three inputs: the vertices and candidate tetrahedra, known ahead of time; the edits for each frame; and the initial set of active tetrahedra.

We apply the method to two matrices.
The first is a proxy graph Laplacian built from tetrahedron edges.
We use it across all meshes because it has no material parameters and shows the behavior of assembly on its own.
The second is a linear elasticity matrix built from \(12\times12\) element blocks.
The update steps are nearly the same for both, so we describe the Laplacian first and then the changes for elasticity in Section~\ref{sec:bridge-to-fem}.

\subsection{Problem Setup}
\label{sec:problem-setup}

Let \(V\in\mathbb{R}^{n\times3}\) hold the fixed reference positions of all candidate vertices, and let \(\mathcal{T}_t\) be the set of active tetrahedra at frame \(t\).
Each tetrahedron \(\tau\) has a set \(E(\tau)\) of six undirected edges.
The \emph{multiplicity} of an edge counts the active tetrahedra that use it:
\begin{equation}
c_t(u,v) = \sum_{\tau \in \mathcal{T}_t} \mathbf{1}[(u,v) \in E(\tau)].
\end{equation}
An edge is active when at least one active tetrahedron uses it:
\begin{equation}
a_t(u,v) = \mathbf{1}[c_t(u,v) > 0].
\end{equation}
The proxy matrix is the graph Laplacian of the active edges,
\begin{equation}
L_t(u,v) =
\begin{cases}
	-a_t(u,v), & u \neq v, \\
	\sum_{w \neq u} a_t(u,w), & u = v,
\end{cases}
\end{equation}
with a small shift to keep it positive definite:
\begin{equation}
A_t = L_t + \varepsilon I.
\end{equation}

The matrix depends only on \(a_t\), but we store \(c_t\).
Without the count, a deleted tetrahedron gives no way to tell whether another active tetrahedron still uses the same edge, short of checking them all.

\subsection{Fixed Matrix Layout}
\label{sec:layout}

Before the simulation starts, we build one compressed sparse row (CSR) layout that holds every edge of every candidate tetrahedron.
We also precompute where each tetrahedron's edges sit in this layout.
During the simulation, only the numerical values change.

An edge that is not active keeps its place in the layout, with a stored value of zero.
The solver therefore sees a matrix with the same values as a rebuilt matrix, though with some extra stored zeros.
Because the layout never changes, no per-frame reconstruction of the operator's sparsity structure is required, and the update writes values directly into the array used by the solver.

\subsection{Incremental Update}
\label{sec:incremental-update}

Each frame provides the deleted tetrahedra \(\Delta_t^-\) and the added tetrahedra \(\Delta_t^+\). Let \(\Delta_t=\Delta_t^-\cup\Delta_t^+\).
The multiplicities change by
\begin{equation}
c_{t+1}(e) = c_t(e) +
\sum_{\tau \in \Delta_t^+} \mathbf{1}[e \in E(\tau)]
-
\sum_{\tau \in \Delta_t^-} \mathbf{1}[e \in E(\tau)].
\end{equation}
We visit only the edges of tetrahedra in \(\Delta_t\).
When a multiplicity moves between zero and a positive value, we write the two off-diagonal entries of that edge and update the diagonal entries at its two endpoints.
Other edges need no matrix write.
Algorithm~\ref{alg:incremental} shows the full loop. In Algorithm~\ref{alg:incremental}, $c$ and $A$ denote the
current counts and matrix, which are updated in place.

\begin{algorithm}[t]
  \caption{Incremental Update loop}
  \label{alg:incremental}
  \begin{algorithmic}[1]
    \Require Vertices $V$, candidate tetrahedra $T$, active set $\mathcal{T}_0$
    \State Build the fixed CSR layout over all of $T$
    \State Compute initial counts $c$ and matrix $A$ from $\mathcal{T}_0$
    \For{$t = 0$ to $F-1$}
      \State Receive edits $\Delta_t^-$ and $\Delta_t^+$
      \State $\Delta_t \gets \Delta_t^- \cup \Delta_t^+$
      \ForAll{$\tau \in \Delta_t$}
        \ForAll{$e = (u,v) \in E(\tau)$}
          \State $c(e) \mathrel{+}= \mathbf{1}[\tau \in \Delta_t^+] - \mathbf{1}[\tau \in \Delta_t^-]$
          \If{$c(e)$ moved between $0$ and a positive value}
            \State Write $A(u,v)$ and $A(v,u)$ in place
            \State Update $A(u,u)$ and $A(v,v)$ in place
          \EndIf
        \EndFor
      \EndFor
      \State Solve $A x_t = b_t$ with preconditioned CG
    \EndFor
  \end{algorithmic}
\end{algorithm}

\subsection{Correctness}
\label{sec:exactness}

\begin{proposition}[Laplacian exactness]
After every frame, the values of the incrementally updated Laplacian equal those of a matrix rebuilt from \(\mathcal{T}_t\).
\end{proposition}

An edit changes the multiplicity only of edges that belong to edited tetrahedra, so all other edges keep the correct activity.
Each edited edge has its correct count, because counts are integers and are updated without error.
Its off-diagonal and diagonal values are written whenever its activity changes, so they match the rebuilt matrix.

\subsection{Element-Assembled Matrices}
\label{sec:bridge-to-fem}

The same idea applies to additive element assembly.
On a fixed set of candidates, many FEM matrices can be written as
\begin{equation}
\widetilde{A}_t = \sum_{\tau \in \mathcal{T}_t} P_{\tau}^{T} K^{(\tau)} P_{\tau} + \varepsilon I,
\end{equation}
where \(K^{(\tau)}\) is the local matrix of element \(\tau\) and \(P_{\tau}\) selects its degrees of freedom from the global vector \cite{sifakis2012fem}.
An edit changes only which terms appear in this sum.
We precompute each \(K^{(\tau)}\) and add or subtract it from the matrix values when \(\tau\) is added or deleted.

Floating-point addition is not exact, so this version matches a rebuild only to within rounding error, and that error can build up over time.
To limit it, we also keep a count of the active elements at each matrix entry.
When a count reaches zero, the entry is reset to its base value, so removed entries leave no residue.
Section~\ref{sec:correctness} measures the remaining error.

\subsection{Compared Methods and Cost}
\label{sec:complexity}

Our main benchmarks compare three policies.
All three write matrices with the same values, up to rounding error on elasticity.

\begin{description}
  \item[Full Rebuild (R)] Recomputes the whole matrix from all active tetrahedra on every frame. It is written in vectorized NumPy and writes into the fixed layout of Section~\ref{sec:layout}.
  \item[Local Recompute (L)] Finds the entries touched by an edit, then recomputes each one by scanning all tetrahedra that use it. It stores no counts or running sums, though it does store the list of tetrahedra at each entry. It also uses the fixed layout.
  \item[Incremental Update (I)] The approach we study, which changes each touched entry using only the edited tetrahedra.
\end{description}

All three policies use NumPy and the same preallocated CSR sparsity pattern, reducing differences due to storage format.
Full Rebuild uses vectorized reassembly, and the solver and timing boundaries are identical across policies.
A separate Python rebuild is used only for correctness validation.

\begin{proposition}[Per-frame assembly cost]
Let \(|\Delta_t|\) be the number of edited tetrahedra, let \(\mathrm{nnz}\) be the number of stored entries, and let \(\kappa\) be the largest number of tetrahedra that share a candidate entry. With the sorting used in our implementation, the cost per frame is:
\begin{align*}
  \text{Full Rebuild} &: O(|\mathcal{T}_t| + \mathrm{nnz}), \\
  \text{Local Recompute} &: O\big(|\Delta_t|\,(\kappa + \log|\Delta_t|)\big), \\
  \text{Incremental Update} &: O(|\Delta_t| \log|\Delta_t|).
\end{align*}
\end{proposition}

The logarithm comes from sorting the touched entries so that repeated entries are combined.
Local Recompute and Incremental Update do not depend on the mesh size when \(\kappa\) is bounded.
These bounds cover updating the values of one operator.
The original dynamic loop in Section~\ref{sec:dynamic} forms a final system from separate operators by visiting every stored entry.
Section~\ref{sec:direct-system} instead treats the fixed-coefficient combined system as the maintained operator.
Practical performance depends on incidence \(\kappa\) and implementation constants, so Section~\ref{sec:results} measures the difference directly.

The fixed layout has costs of its own.
Setup must build the layout and index maps, memory must hold them, and stored zeros add work to matrix-vector products.
Section~\ref{sec:setup-memory} reports setup and memory.

\subsection{Connectivity}

Our pipeline also tracks which tetrahedra are connected.
We use union-find with path compression and union by rank, rebuilt from the active set every five frames.
We chose this over an incremental disjoint-set structure \citep{tarjan1975setunion}, which does not support deletion, and over fully dynamic connectivity \citep{holm2001dynamic}, which is much harder to implement.
Between rebuilds the answers can be out of date, as Section~\ref{sec:correctness} shows.
Connectivity is not used to build the matrix, so this does not affect matrix accuracy.
The reverse also holds: a correct matrix says nothing about connectivity, and a simulator that relies on these answers would need fresher ones.

\section{Implementation}

We implemented the method in Python with NumPy and SciPy.
In the main benchmarks, all three policies write into one precomputed CSR layout and pass it to SciPy's conjugate gradient solver.
Edits are applied as array operations, not element by element.

The method replaces the assembly step of a solver when the candidate pool and element contributions can be prepared in advance.
It does not update the preconditioner or define how simulation states should change when elements become active.

\section{Experimental Setup}

\subsection{Models and Scenarios}

We use three objects from \citet{cutler2004simplification}: Bunny (10k, 50k, and 460k elements), Gargoyle (50k), and Hand (100k).
The direct-system experiment additionally uses Bunny 2k.
The main benchmarks use three scripted edit scenarios.
In \textbf{fracture}, tetrahedra in a slab through the middle of the object are deleted over time.
In \textbf{refinement}, a random 3\% of tetrahedra are each replaced by four children that meet at the parent's centroid.
In \textbf{merge}, the simulation starts with a slab removed and adds its tetrahedra back over time.
These are scripted edits that remove, subdivide, and reactivate tetrahedra.
They test matrix updates and do not cover everything a full fracture or merge simulation must do, such as finding cracks or handling contact.

\subsection{Experiments}

Unless noted, each configuration runs all three policies for 50 seeds.

\begin{description}
  \item[Proxy Laplacian] Bunny 50k, Gargoyle, Hand, and Bunny 460k, with all three scenarios. Runs have 12 frames, except Bunny 460k with 8.
  \item[Linear elasticity] Bunny 10k, all three scenarios, 4 frames, with Young's modulus 1 and Poisson's ratio 0.3.
  \item[Synthetic dynamic elasticity] An implicit Euler loop on Bunny 10k for fracture and merge, 4 frames.
  \item[Direct system assembly] Bunny 2k, 10k, and 50k, with fixed batches of eight tetrahedra, 30 trials per configuration, and 200 measured frames after 20 warm-up frames per trial.
  \item[Refinement with activated reservations] A four-step Bunny 50k replay with refinement seed 0 and 12 paired timing trials, comparing Direct Incremental with bulk initialization against compact reassembly.
  \item[Slab width] Proxy Laplacian fracture on Bunny 10k with slab half-widths 0.01, 0.03, and 0.08, 8 frames.
  \item[Drift] One 5,000-frame elasticity run on Bunny 10k with mixed fracture, merge, and refinement edits, for each of three seeds.
\end{description}

We use conjugate gradients (CG) with a diagonal preconditioner and a requested relative tolerance of \(10^{-8}\).
The iteration cap is 2,000 for the proxy runs and 4,000 for elasticity.
Each solve starts from zero, and the matrix shift uses \(\varepsilon=10^{-6}\).
The reported timing environment uses an absolute tolerance of zero.
Section~\ref{sec:correctness} reports convergence checks separately from matrix agreement.
The right-hand side in the Laplacian and static elasticity runs is a random vector, fixed for each run.
These runs measure the cost of the solver and do not model a physical load.

\subsection{Timing}

\emph{Assembly time} is all time spent making the matrix ready for the solver, including any matrix construction.
Benchmarked \emph{frame time} includes assembly, the linear solve, connectivity, and one residual check.
Matrix-parity checks are not part of the timing runs.
A separate pass replays the same edits and compares each matrix against a plain Python rebuild, which serves as the reference.
The direct preparation experiment and the refinement replay use the timing boundaries given in Sections~\ref{sec:direct-system} and~\ref{sec:refinement-reservations}.

All reported timing measurements were collected on the lab machine, one process at a time, with numerical libraries limited to one thread: an AMD EPYC 75F3 with 504\,GB of RAM, running Linux, Python 3.13.5, NumPy 2.2.6, and SciPy 1.15.2.
The machine is shared with other users.
For the original benchmark configurations, we report means and standard deviations over seeds, and ratios of means.
The refinement replay instead measures timing variation across repeated trials of one edit sequence.
The later diagnostic replays check numerical behavior and convergence; they do not replace these timing measurements.

\section{Results}
\label{sec:results}

\subsection{Assembly Across Models}
\label{sec:multi-dataset}

\begin{table*}[t]
  \caption{Proxy Laplacian results across all models. Mean\,$\pm$\,std over 50 seeds. R\,=\,Full Rebuild, L\,=\,Local Recompute, I\,=\,Incremental Update. Bold marks the lowest mean in each group; see Section~\ref{sec:significance} for which differences are significant.}
  \label{tab:laplacian}
  \resizebox{\textwidth}{!}{%
  \begin{tabular}{llrrrrrr}
    \toprule
    & & \multicolumn{3}{c}{Assembly time (ms)} & \multicolumn{3}{c}{Frame time (ms)} \\
    \cmidrule(lr){3-5}\cmidrule(lr){6-8}
    Model & Scenario & R & L & I (ours) & R & L & I (ours) \\
    \midrule
Bunny 50k & Fracture & $1.57\pm0.01$ & $0.124\pm0.003$ & $\mathbf{0.115\pm0.003}$ & $56.56\pm0.33$ & $\mathbf{54.88\pm0.36}$ & $54.98\pm0.59$ \\
 & Refinement & $1.73\pm0.02$ & $0.274\pm0.006$ & $\mathbf{0.267\pm0.005}$ & $60.58\pm0.28$ & $59.15\pm0.34$ & $\mathbf{59.02\pm0.25}$ \\
 & Merge & $1.57\pm0.02$ & $0.122\pm0.003$ & $\mathbf{0.113\pm0.003}$ & $56.47\pm0.27$ & $\mathbf{54.85\pm0.38}$ & $54.89\pm0.35$ \\
\midrule
Gargoyle 50k & Fracture & $1.65\pm0.01$ & $0.143\pm0.003$ & $\mathbf{0.127\pm0.003}$ & $75.99\pm0.39$ & $74.32\pm0.90$ & $\mathbf{74.26\pm0.57}$ \\
 & Refinement & $1.81\pm0.01$ & $\mathbf{0.277\pm0.006}$ & $0.285\pm0.006$ & $80.73\pm0.37$ & $\mathbf{79.15\pm0.34}$ & $79.19\pm0.48$ \\
 & Merge & $1.65\pm0.01$ & $0.140\pm0.003$ & $\mathbf{0.125\pm0.003}$ & $75.93\pm0.31$ & $\mathbf{74.18\pm0.54}$ & $74.22\pm0.54$ \\
\midrule
Hand 100k & Fracture & $3.41\pm0.04$ & $0.237\pm0.010$ & $\mathbf{0.181\pm0.006}$ & $334.8\pm2.3$ & $\mathbf{331.1\pm2.4}$ & $331.9\pm2.5$ \\
 & Refinement & $3.77\pm0.05$ & $0.532\pm0.013$ & $\mathbf{0.450\pm0.009}$ & $356.9\pm3.1$ & $354.0\pm3.3$ & $\mathbf{353.3\pm3.6}$ \\
 & Merge & $3.41\pm0.05$ & $0.231\pm0.008$ & $\mathbf{0.180\pm0.005}$ & $334.5\pm2.2$ & $\mathbf{330.8\pm2.6}$ & $331.2\pm2.9$ \\
\midrule
Bunny 460k & Fracture & $16.33\pm0.25$ & $0.963\pm0.023$ & $\mathbf{0.601\pm0.014}$ & $1,017.7\pm11.7$ & $1,007.2\pm13.3$ & $\mathbf{1,006.0\pm13.5}$ \\
 & Refinement & $17.95\pm0.21$ & $3.26\pm0.06$ & $\mathbf{2.27\pm0.04}$ & $1,103.0\pm17.1$ & $1,092.5\pm16.4$ & $\mathbf{1,079.5\pm12.8}$ \\
 & Merge & $16.32\pm0.17$ & $0.946\pm0.023$ & $\mathbf{0.600\pm0.027}$ & $1,018.9\pm14.7$ & $1,005.5\pm14.2$ & $\mathbf{1,001.8\pm12.5}$ \\
    \bottomrule
  \end{tabular}}
\end{table*}

\begin{figure}[t]
  \centering
  \includegraphics[width=\linewidth]{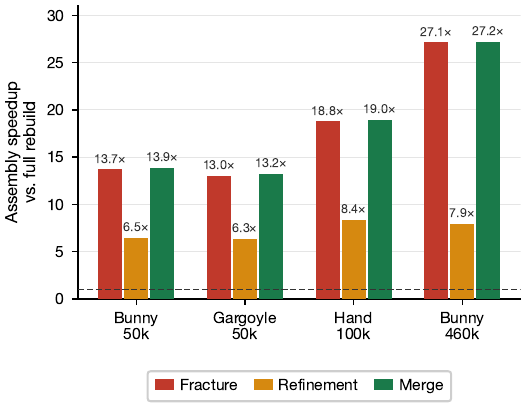}
  \caption{Assembly speedup of Incremental Update over Full Rebuild. The dashed line marks no speedup. For fracture and merge, the speedup grows with mesh size.}
  \Description{Bar chart. For Bunny 50k, Gargoyle 50k, Hand 100k, and Bunny 460k, fracture and merge speedups are about 14, 13, 19, and 27. Refinement speedups are about 6.5, 6.3, 8.4, and 7.9.}
  \label{fig:cross-dataset}
\end{figure}

Table~\ref{tab:laplacian} gives the full proxy Laplacian results, and Figure~\ref{fig:cross-dataset} shows the gain over Full Rebuild.
Incremental Update is faster than R in every configuration.
The gain is 13 to 27 times for fracture and merge, and 6 to 8 times for refinement, where each frame edits more elements.
For fracture and merge, the gain also grows with mesh size, from about 14 times on Bunny 50k to about 27 times on Bunny 460k.

\begin{figure}[t]
  \centering
  \includegraphics[width=\linewidth]{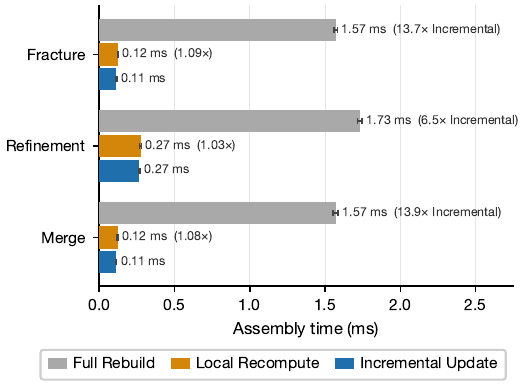}
  \caption{Assembly time on Bunny 50k, shown as mean\,$\pm$\,std. Values in parentheses are ratios to Incremental Update.}
  \Description{Horizontal bar chart. Full Rebuild takes 1.57 to 1.73 milliseconds. Local Recompute and Incremental Update take about 0.11 to 0.27 milliseconds and are close to each other.}
  \label{fig:update-cost}
\end{figure}

Figure~\ref{fig:update-cost} compares the three policies on Bunny 50k.
Incremental Update and Local Recompute are close on this mesh.
The difference between them grows on larger meshes, reaching about 1.6 times for Bunny 460k fracture and merge.

\subsection{Cost Follows Edit Size}
\label{sec:scaling}

\begin{table}[t]
  \caption{Assembly time grows with the number of matrix edges that change. Rows are sorted by changed edges. Edited tetrahedra and changed edges are per-frame means. Merge matches fracture and is omitted.}
  \label{tab:edit-size}
  \resizebox{\linewidth}{!}{%
  \begin{tabular}{llrrrrrr}
    \toprule
    & & & & & \multicolumn{3}{c}{Assembly (ms)} \\
    \cmidrule(lr){6-8}
    Model & Scenario & Active & Edited & Changed & R & L & I \\
    & & tets & tets & edges & & & \\
    \midrule
Bunny 50k & Fracture & 50k & 48 & 38 & 1.57 & 0.12 & 0.11 \\
Gargoyle 50k & Fracture & 49k & 81 & 67 & 1.65 & 0.14 & 0.13 \\
Hand 100k & Fracture & 99k & 179 & 145 & 3.41 & 0.24 & 0.18 \\
Bunny 50k & Refinement & 52k & 625 & 501 & 1.73 & 0.27 & 0.27 \\
Gargoyle 50k & Refinement & 52k & 625 & 503 & 1.81 & 0.28 & 0.29 \\
Hand 100k & Refinement & 105k & 1,250 & 1,002 & 3.77 & 0.53 & 0.45 \\
Bunny 460k & Fracture & 456k & 1,126 & 1,357 & 16.33 & 0.96 & 0.60 \\
Bunny 460k & Refinement & 484k & 8,639 & 6,915 & 17.95 & 3.26 & 2.27 \\
    \bottomrule
  \end{tabular}}
\end{table}

Table~\ref{tab:edit-size} shows why the gain varies.
The time for R follows the mesh: it stays near 1.6--1.8\,ms on both 50k meshes and reaches 16--18\,ms on Bunny 460k, whatever the edit.
The time for Incremental Update follows the edit: it rises steadily with the number of changed edges, from 0.11\,ms at 38 edges to 2.27\,ms at 6,915.
From Bunny 50k fracture to Bunny 460k fracture, R takes about 10 times longer, while Incremental Update takes about 5 times longer on a frame with about 23 times more edited tetrahedra.

Very small edits show a fixed minimum cost.
In our slab width test on Bunny 10k, widening the slab raises the edit from about 4 to 29 tetrahedra per frame, while Incremental Update assembly rises only from 0.082\,ms to 0.101\,ms.
Local Recompute behaves the same way (0.080\,ms to 0.105\,ms).
R stays at 0.34\,ms.

\subsection{Where Frame Time Goes}
\label{sec:frame-time}

\begin{figure}[t]
  \centering
  \includegraphics[width=\linewidth]{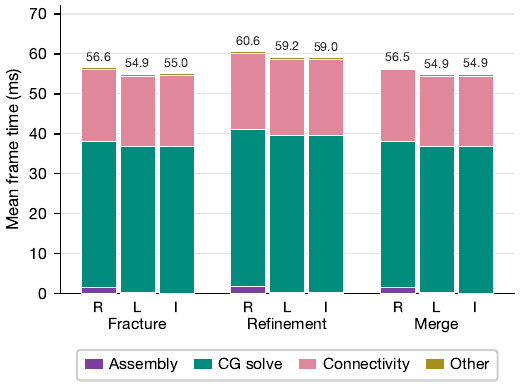}
  \caption{Frame time on Bunny 50k, split by component and labeled with the total in ms. Assembly is the thin bottom layer. Connectivity is averaged over frames; in practice it is a rebuild of about 100\,ms on 2 of every 12 frames.}
  \Description{Stacked bar chart. All policies total about 55 to 61 milliseconds. CG takes about 37 to 39 milliseconds, connectivity about 17 to 19 milliseconds, and assembly at most about 1.7 milliseconds.}
  \label{fig:bunny-frame-time}
\end{figure}

A faster assembly step helps a frame only as much as assembly matters to that frame.
Figure~\ref{fig:bunny-frame-time} splits frame time into its parts.
In the proxy runs, the solve takes 65--89\% of each frame for all three policies, and connectivity rebuilds take most of the rest.
Assembly is at most 0.5\% of an Incremental Update frame and 1--3\% of an R frame.

As a result, frames with Incremental Update are only 0.9--2.8\% shorter than with R on the Laplacian.
These are small reductions in mean frame time compared with the much larger assembly savings.

\subsection{Linear Elasticity}
\label{sec:stiffness-transfer}

\begin{table*}[t]
  \caption{Linear elasticity results on Bunny 10k. Mean\,$\pm$\,std over 50 seeds. R\,=\,Full Rebuild, L\,=\,Local Recompute, I\,=\,Incremental Update. Bold marks the lowest mean in each group; see Section~\ref{sec:significance} for which differences are significant.}
  \label{tab:elasticity}
  \resizebox{\textwidth}{!}{%
  \begin{tabular}{llrrrrrr}
    \toprule
    & & \multicolumn{3}{c}{Assembly time (ms)} & \multicolumn{3}{c}{Frame time (ms)} \\
    \cmidrule(lr){3-5}\cmidrule(lr){6-8}
    Model & Scenario & R & L & I (ours) & R & L & I (ours) \\
    \midrule
Bunny 10k & Fracture & $4.52\pm0.33$ & $0.336\pm0.023$ & $\mathbf{0.267\pm0.026}$ & $152.4\pm6.0$ & $\mathbf{148.2\pm6.0}$ & $148.4\pm6.4$ \\
 & Refinement & $4.77\pm0.19$ & $\mathbf{2.07\pm0.06}$ & $2.10\pm0.03$ & $169.3\pm5.9$ & $166.2\pm5.4$ & $\mathbf{166.0\pm5.4}$ \\
 & Merge & $4.45\pm0.15$ & $0.334\pm0.010$ & $\mathbf{0.257\pm0.010}$ & $151.1\pm6.7$ & $\mathbf{146.9\pm6.0}$ & $147.4\pm6.3$ \\
    \bottomrule
  \end{tabular}}
\end{table*}

Table~\ref{tab:elasticity} gives the results for linear elasticity, where each element contributes a \(12\times12\) block containing up to 144 local entries.
For fracture and merge, Incremental Update assembles about 17 times faster than R and about 1.3 times faster than Local Recompute.

Refinement is the weak case.
It edits about 16 times more tetrahedra per frame than fracture (374 against 23), and assembly takes 2.1\,ms.
The gain over R falls to 2.3 times, and the mean assembly times of the two local policies are close.

Frame time is dominated by the solve, which takes 97--99\% of each frame for all three policies.
Frames with Incremental Update are 2.0--2.6\% shorter than with R.

\subsection{Synthetic Dynamic Elasticity}
\label{sec:dynamic}

We also compare the policies in a synthetic implicit Euler loop on Bunny 10k.
Each frame solves
\begin{equation}
(M_t + h^2 K_t)\,u_{t+1} = M_t(u_t + h v_t) + h^2 f,
\label{eq:dynamic}
\end{equation}
where \(u_t\) is a state vector with three components per candidate vertex and \(v_t\) is its rate of change.
Both start at zero, and after each solve we set \(v_{t+1}=(u_{t+1}-u_t)/h\).
Reference vertex positions remain fixed.
We use \(h=0.033\), density \(\rho=1\), Young's modulus 1, and Poisson's ratio 0.3.
These parameters define a synthetic workload; we do not assign a physical unit system.

The diagonal matrix \(M_t\) stores lumped mass.
Each active tetrahedron contributes \(\rho\,\mathrm{volume}/4\) to each coordinate's mass at each of its four vertices.
The matrix \(K_t\) contains the assembled linear-elasticity stiffness and the shift \(\varepsilon I\), with \(\varepsilon=10^{-6}\).
The combined system therefore includes \(h^2\varepsilon I\).
The load \(f\) is a seed-dependent vector of independent standard normal samples scaled by 0.01, drawn once and held fixed.
It acts on every candidate DOF, and no boundary constraints are imposed.

Each frame applies the edits, updates stiffness and mass, forms the right-hand side, solves, and advances the state.
DOFs with no active incident tetrahedra retain their load and continue to evolve under the diagonal shift.
They can acquire large values, which are carried into the active system if elements later use those DOFs.
Activation does not reset or interpolate the state.
This loop tests assembly within an evolving algebraic system; it is not a validated physical fracture or activation model.

Solves take about 17--18\,ms in this workload, compared with about 150\,ms in the static elasticity study.
Assembly therefore accounts for a larger share of the frame: about 20\% for Full Rebuild.
Table~\ref{tab:dynamic} gives the results.
Mean frame time falls from 23.2 to 18.9\,ms for fracture and from 22.5 to 18.1\,ms for merge, a reduction of 18--20\%.
Local Recompute reaches nearly the same frame times, at 19.0 and 18.1\,ms.

Assembly is about 7.5 times faster with Incremental Update than with Full Rebuild.
The gain is smaller than in the static fracture and merge tests because forming \(M_t+h^2K_t\) still visits every stored entry.
This combination step takes 46--47\% of Incremental Update's assembly time.
We include it in assembly time for every policy.

\begin{table}[t]
  \caption{Dynamic elasticity on Bunny 10k. Mean\,$\pm$\,std over 50 seeds, in ms. The last column gives the estimated frames needed for Incremental Update to repay its extra setup over Full Rebuild through faster assembly.}
  \label{tab:dynamic}
  \resizebox{\linewidth}{!}{%
  \begin{tabular}{llrrrr}
    \toprule
    Scenario & Policy & Assembly & Frame & Setup & Est. frames to repay \\
    \midrule
Fracture & R & $4.71\pm0.25$ & $23.21\pm0.38$ & 433.6 & -- \\
 & L & $0.731\pm0.044$ & $19.01\pm0.26$ & 760.8 & -- \\
 & I & $0.632\pm0.033$ & $18.92\pm0.29$ & 744.9 & 76 \\
Merge & R & $4.80\pm0.52$ & $22.53\pm1.36$ & 449.2 & -- \\
 & L & $0.696\pm0.029$ & $18.14\pm0.77$ & 774.1 & -- \\
 & I & $0.639\pm0.052$ & $18.12\pm0.83$ & 757.1 & 74 \\
    \bottomrule
  \end{tabular}}
\end{table}

The setup times in Table~\ref{tab:dynamic} exclude the early construction of the common mass cache and initial mass vector.
They are therefore not complete setup totals and should not be compared directly with the inclusive setup times in Section~\ref{sec:refinement-reservations}.

\subsection{Direct Time-Step Matrix Assembly}
\label{sec:direct-system}

The current dynamic path updates \(K_t\) and lumped mass locally, then copies the CSR matrix and forms \(A_t=M_t+h^2K_t\) with a pass over its stored values.
For fixed \(h\), reference geometry, material, and density, the direct paths cache \(A_e=M_e+h^2K_e\).
Direct Incremental applies signed element contributions; Direct Local Recompute rescans active contributors to affected entries.
Both retain the candidate pool, CSR layout, and global diagonal shift \(h^2\varepsilon\), and maintain lumped mass for the right-hand side.

The preparation timer runs from the supplied edit batch until \(A_t\) and the current mass vector are ready.
It includes edit validation, matrix and mass updates, zero-support resets, and any copying, combination, or materialization.
All paths in this experiment share a mass-update helper that checks edit IDs and resets mass to zero when the last supporting element is removed.
The main dynamic loop in Section~\ref{sec:dynamic} uses its original mass-update routine.
Setup, reference reconstruction, right-hand-side formation, preconditioning, and solving are excluded.

\begin{figure}[t]
  \centering
  \includegraphics[width=\linewidth]{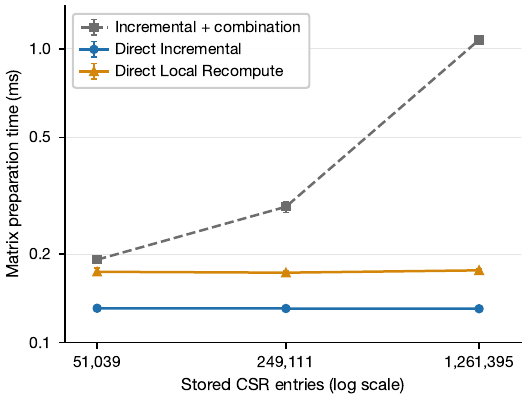}
  \caption{Matrix preparation time for fixed-step dynamic elasticity on Bunny 2k, 10k, and 50k. Each edit batch changes eight vertex-disjoint tetrahedra and touches 1,152 stored CSR entries. Points and error bars show the mean and sample standard deviation of 30 trial means per configuration (three seeds, ten repetitions each). Both axes are logarithmic. Timing includes matrix and lumped mass updates and any required copying or combination. Right-hand-side formation and solving are excluded.}
  \Description{Three curves against actual stored CSR entry counts 51,039, 249,111, and 1,261,395. Incremental plus combination increases from 0.192 to 1.074 milliseconds. Direct Incremental stays near 0.131 milliseconds, and Direct Local Recompute near 0.175 milliseconds. Error bars show standard deviations across 30 trials.}
  \label{fig:direct-system}
\end{figure}

Figure~\ref{fig:direct-system} shows the separate preparation experiment.
As stored entries grow from 51,039 to 1,261,395, Incremental + combination rises from 0.192 to 1.074\,ms.
Direct Incremental stays near 0.131\,ms and Direct Local Recompute at 0.173--0.176\,ms: both are nearly flat over this tested range at fixed edit size.
On Bunny 50k, Direct Incremental reduces preparation time by 0.943\,ms, an 8.23\(\times\) speedup over the Incremental + combination path in this experiment, including its CSR copy and combination pass.
This is a preparation-only comparison, not an additional full-frame speedup; right-hand-side and solver work remain global.
Untimed Bunny 50k checks cover deletion, restoration, and trial-end states for all paths in all 30 trials (270 comparisons).
Independent reconstruction of \(M_t\) and \(K_t\), with the same active elements and shift, gives maximum matrix and mass-vector differences of \(1.1\times10^{-19}\) and \(4.2\times10^{-22}\), below the \(10^{-10}\) tolerance.

\subsection{Refinement with Activated Reservations}
\label{sec:refinement-reservations}

We next test whether direct assembly savings persist when refinement activates reserved entries and the comparison includes setup and matrix application.
The four-step Bunny 50k replay removes 375 parent tetrahedra and activates 1,500 centroid-refinement children per step.
Both paths use the same pool of 56,000 candidate tetrahedra, 36,603 reserved degrees of freedom, cached element contributions, and material and time-step parameters.
Each step activates 29,250 previously unused structural entries; all 117,000 initially unused entries are active by the final step.

Direct Incremental maintains the combined operator from Section~\ref{sec:direct-system} on its reserved layout.
It initializes numerical values and integer support counts by bulk accumulation over the initial active elements, and then uses the unchanged incremental edit operations.
Compact rebuild reassembles all active contributions and constructs a CSR matrix for the current structural support after each edit.
Unlike Full Rebuild in the main benchmarks, it does not retain slots used only by inactive candidates.
Every active pattern occurs once in this replay, so we do not use a cache of previously assembled compact patterns.

This is an edited-system replay.
An independently assembled reference trajectory generates the right-hand sides, which are saved and used unchanged by both timed paths.
Both paths start each CG solve from zero and do not advance their own simulation states during timing.
The reference trajectory uses the same load and inactive-state rules as Section~\ref{sec:dynamic}, with no centroid interpolation on activation.
Newly active DOFs can therefore carry large stored states into the reference trajectory.
The shared inputs provide a controlled algebraic comparison, but they do not validate a physical activation model.
Each of 12 paired timing trials restarts from the initial active set and assembled matrices, with path order varied after warm-up.
Setup includes the element caches, layouts, index maps, mass metadata, and maintained initial state, with shared construction costs charged once to each standalone path.
The four edited steps include matrix and mass preparation, Jacobi setup, and CG.
Right-hand-side generation and correctness checks are excluded.

\begin{table}[t]
  \centering
  \caption{Refinement with activated reservations on Bunny 50k. Means over 12 paired trials of one four-step replay. Direct Incremental uses bulk initialization. Preparation, Jacobi, and CG times sum all four edits; the last row includes one-time setup.}
  \label{tab:refinement-reservations}
  \small
  \setlength{\tabcolsep}{4pt}
  \begin{tabular}{@{}lrr@{}}
    \toprule
    Stage & \shortstack{Direct\\Incremental} & \shortstack{Compact\\rebuild} \\
    \midrule
    Setup (s) & 2.109 & 2.112 \\
    Preparation (ms) & 63.66 & 147.13 \\
    Jacobi setup (ms) & 4.97 & 3.73 \\
    CG (ms) & 518.71 & 506.83 \\
    Four edited steps (ms) & 587.34 & 657.69 \\
    \midrule
    Setup + four steps (s) & \textbf{2.696} & 2.770 \\
    \bottomrule
  \end{tabular}
\end{table}

Table~\ref{tab:refinement-reservations} shows comparable setup costs.
Direct Incremental saves 83.47\,ms in preparation across the four edits, exceeding its additional 13.12\,ms in Jacobi setup and CG.
Including setup, it saves 73.8\,ms, or approximately 2.7\%, with a paired 95\% bootstrap interval of 71.2--77.2\,ms.
Retained assembly buffers occupy 159.91\,MiB for Direct and 160.05\,MiB for compact rebuild, including caches; these figures exclude solver workspace and are not process memory measurements.

Untimed checks against independent reconstruction cover the initial state and every edit, before timing and on the first and last trials.
Maximum differences do not exceed \(1.08\times10^{-19}\) for the matrix, \(5.08\times10^{-21}\) for mass, and \(4.34\times10^{-19}\) for the right-hand side.
Activity, support counts, diagonal shifts, and exact zeros outside active support also pass validation, including restoration and last-contributor removal.
Both reported paths converge in 172, 174, 180, and 182 CG iterations across the four steps, with relative residuals below \(10^{-8}\).
This replay shows that direct assembly savings can survive initialization and matrix application when reserved entries become active.
The result is specific to this sequence; unused stored entries can still increase application costs.

\subsection{Statistical Analysis}
\label{sec:significance}

The main comparison contains 20 configurations: 12 proxy cases, three static elasticity cases, two synthetic dynamic cases, and three slab widths.
We compare Incremental Update with each other policy on assembly and frame time, pairing runs by seed.
This gives 80 comparisons.
We apply a paired t-test and a Wilcoxon signed-rank test, with Holm correction over the 80 comparisons for each test.
A difference is counted as significant only when both adjusted \(p\)-values are below 0.05.
The separate direct-system and refinement experiments are excluded from this family of tests.

Incremental Update is significantly faster than Full Rebuild in all 20 configurations, for both assembly and frame time.
Table~\ref{tab:significance} compares Incremental Update with Local Recompute.
Incremental Update has lower assembly time in 16 configurations, while Local Recompute is faster in two; the remaining two show no significant difference.
The frame-time differences are smaller: Incremental Update is faster in two configurations, Local Recompute in one, and no significant difference is detected in the remaining 17.
Thus, lower assembly time often does not produce a detectable frame-time improvement over Local Recompute.

\begin{table}[t]
  \centering
  \caption{Incremental Update (I) against Local Recompute (L) across 20 configurations. Counts show significant differences after Holm correction over the 80 comparisons, requiring both statistical tests to pass.}
  \label{tab:significance}
  \small
  \setlength{\tabcolsep}{4pt}
  \begin{tabular}{@{}lrrr@{}}
    \toprule
    Metric & I faster & L faster & \shortstack{No significant\\difference detected} \\
    \midrule
    Assembly time & 16 & 2 & 2 \\
    Frame time & 2 & 1 & 17 \\
    \bottomrule
  \end{tabular}
\end{table}

For the refinement replay, we use a paired bootstrap interval from 20,000 resamples of whole timing trials.
Each resample keeps the four-step sequence together.
This interval describes timing variation for that one sequence, not variation across different meshes or edit schedules.

\subsection{Setup and Memory}
\label{sec:setup-memory}

\begin{table}[t]
  \caption{One-time setup cost and retained memory, measured with \texttt{tracemalloc}.}
  \label{tab:cost}
  \resizebox{\linewidth}{!}{%
  \begin{tabular}{lrrr}
    \toprule
    Case & R & L & I \\
    \midrule
Laplacian, Bunny 50k: setup (ms) & 18.1 & 40.3 & 27.5 \\
Laplacian, Bunny 50k: memory (MB) & 4.92 & 9.73 & 7.13 \\
Elasticity, Bunny 10k: setup (ms) & 432.3 & 503.8 & 490.3 \\
Elasticity, Bunny 10k: memory (MB) & 24.89 & 50.82 & 27.91 \\
    \bottomrule
  \end{tabular}}
\end{table}

Table~\ref{tab:cost} shows the cost of preparing for fast updates in the original benchmark implementations.
Incremental Update uses less setup time and memory than Local Recompute, because it does not store the list of tetrahedra at each entry.
It uses more of both than R.

We estimate that the extra setup over R is repaid through faster assembly after about 6--7 frames on the Laplacian, 14--22 frames on elasticity, and about 75 frames in the dynamic loop.
These estimates use the recorded setup differences and assembly savings in the original implementations.
The dynamic estimate excludes the common mass preparation described in Section~\ref{sec:dynamic}.
Section~\ref{sec:refinement-reservations} separately includes setup, preparation, preconditioning, and solving for Direct Incremental with bulk initialization against compact rebuild.

\subsection{Correctness and Solver Checks}
\label{sec:correctness}

Table~\ref{tab:parity} summarizes the separate matrix-parity checks for the retained benchmarks.
No matrix-parity check failed.
The Laplacian matched the reference exactly, and elasticity stayed within \(1.1\times10^{-16}\) on these short runs.
The pattern of nonzero entries also matched in the Laplacian and elasticity runs, not counting stored zeros.

\begin{table}[t]
  \caption{Correctness checks against the Python reference, over all three policies. A failure means any difference on the Laplacian, or a difference above \(10^{-10}\) on elasticity.}
  \label{tab:parity}
  \begin{tabular}{lrrr}
    \toprule
    Experiment & Checks & Failures & Largest difference \\
    \midrule
    Proxy Laplacian & 19,800 & 0 & 0 \\
    Elasticity & 1,800 & 0 & \(1.1\times10^{-16}\) \\
    Dynamic fracture & 600 & 0 & \(5.4\times10^{-20}\) \\
    Dynamic merge & 600 & 0 & \(5.4\times10^{-20}\) \\
    \bottomrule
  \end{tabular}
\end{table}

\begin{figure}[t]
  \centering
  \includegraphics[width=\linewidth]{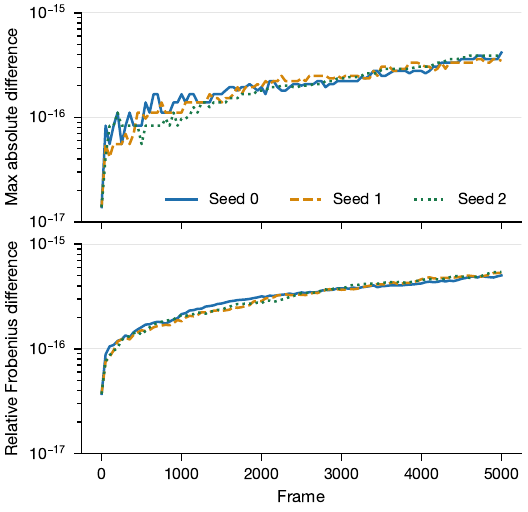}
  \caption{Rounding error of incremental elasticity against the Python reference over 5,000 frames, for three seeds. The error grows slowly and stays near machine precision.}
  \Description{Two line plots on log axes. The maximum absolute difference and relative Frobenius difference rise from about 1e-17 to about 5e-16 over 5000 frames.}
  \label{fig:drift}
\end{figure}

Over longer runs, rounding error does build up (Figure~\ref{fig:drift}).
Across 5,000 frames, the largest entry difference rises from \(1.4\times10^{-17}\) to \(4.2\times10^{-16}\).
The pattern of entries never differed.
The error stays far below our \(10^{-10}\) check, but it does grow.
A long simulation could remove it with an occasional rebuild.

Elasticity also shows a small effect on the solver.
Total CG iterations matched across all three policies in only 11 of 150 seed and scenario pairs, with a median spread of 3 iterations out of about 3,400 and a largest spread of 159.
On the Laplacian, iteration counts matched in all 600 pairs.
We suspect that CG is reacting to rounding-level differences in the matrix, but we did not test this.
For this reason, small differences in elasticity frame time cannot be credited to assembly alone, and we report assembly and frame time separately.

We also replayed the static and synthetic dynamic elasticity workloads with additional solver diagnostics, using all 50 seeds and three policies.
Across the 3,000 solves in these retained configurations---1,800 static and 1,200 dynamic---every solve reported success and passed an independently recomputed residual check at the requested tolerance.
No solve reached the iteration cap or produced nonfinite values.
The historical elasticity logs did not retain solver return codes or right-hand-side norms, so these replay checks provide new evidence rather than recover the missing historical records.
The diagnostic runs did not reproduce historical iteration counts exactly, so their timings are not used as replacements for the original measurements.

In the saved proxy and slab-width runs, all 23,400 solves reported success and none reached its iteration cap.
In the refinement comparison, all 96 reported solves have saved success codes and independently recomputed relative residuals below \(10^{-8}\).
These solver checks are separate from matrix-parity checks: agreement between assembled matrices does not by itself establish convergence.

A separate check of the main dynamic workloads found a maximum mass-vector difference of \(2.03\times10^{-20}\) from independent reconstruction.
Residual mass on unsupported DOFs was at most \(1.22\times10^{-13}\) of the diagonal shift \(h^2\varepsilon\).
All 1,200 checked edit batches had consistent active sets and support counts.
Relative \(L_2\) differences between the policies' supported states remained below \(8.07\times10^{-9}\).
These checks found no consequential mass-maintenance error on the tested dynamic workloads.

Connectivity disagreed with a breadth-first search in 28 of 1,800 checks (1.6\%).
All disagreements were in refinement and merge, with the highest rate at 6\% on Bunny 460k refinement.
Fracture had none.
This comes from rebuilding union-find only every five frames, and it does not affect the matrix.

\section{Discussion}

\subsection{When to Use Incremental Update}

The method fits best when:
\begin{itemize}
  \item the candidate elements can be allocated in advance;
  \item each frame edits a small part of the mesh, as in fracture and merge;
  \item the mesh is large, since rebuild cost grows with the mesh;
  \item assembly is a visible part of the frame, as it is when solves are cheap.
\end{itemize}

The method offers less when the solve dominates the frame or when edits are large.
For short sequences, setup costs should also be included in the comparison.
Local Recompute is also a strong choice.
It reaches nearly the same frame time, and it avoids history-dependent accumulation of rounding error, at the cost of more memory and setup time.
In the tested large-mesh cases, Incremental Update reduces assembly time compared with Local Recompute; it also uses less retained memory in the reported measurements.
For a fixed-coefficient additive time-step matrix, the direct variant also removes the original path's global combination pass (Section~\ref{sec:direct-system}); this does not make right-hand-side formation, preconditioning, or solving local.
In the refinement replay, bulk initialization makes setup comparable to compact rebuild, and preparation savings exceed the additional matrix application cost (Section~\ref{sec:refinement-reservations}).
The gain including setup is modest, so preparation speedups alone are insufficient to choose an assembly path.

\subsection{Limitations}

Our results describe these Python, NumPy, and SciPy implementations, the tested meshes and edit sizes, and two operators.
A compiled implementation could change the relative costs.
The lab machine is shared, and we retain completed timing runs rather than discard slow ones.
We use a diagonal preconditioner and start each solve from zero; other solver choices could change how much of the frame is spent solving.

The loads and edit schedules are synthetic.
In the evolving loop, unsupported DOFs retain their loads and states, and activation provides no state transfer.
Large stored states can enter the active system during merge and refinement.
The refinement replay gives both paths identical saved right-hand sides, so it controls the inputs for the comparison, but it does not establish realistic activation behavior.
The reported frame-time gains should therefore be read as results for these algebraic workloads, not predictions for a complete physical simulator.

Most timing sequences are short, and the 12-frame proxy runs include only two connectivity rebuilds.
The refinement replay uses one mesh and one edit sequence; repeated trials measure timing variation on that sequence, not performance across other workloads.
The main dynamic mass updater assumes valid edit schedules, which held in the checked runs; the direct preparation paths also validate edit IDs and reset unsupported mass entries.

The method requires a fixed candidate pool and does not apply as described to unrestricted remeshing.
It does not maintain preconditioners or factorizations.
The elasticity matrices use fixed reference geometry and material parameters, so element contributions can be cached.
Nonlinear materials or changing reference geometry would require additional element updates.

\section{Conclusion}

We evaluated incremental matrix assembly for prescribed tetrahedral edits on a fixed candidate pool.
In our tests, assembly is 6--27 times faster than vectorized full reassembly for a proxy Laplacian and 2--17 times faster for linear elasticity.
The update work follows the edited elements rather than the whole mesh.
Frame-time savings are smaller: 1--3\% in the solve-dominated benchmarks and 18--20\% in the synthetic dynamic elasticity loop with shorter solves.
Local Recompute often reaches similar frame times.
The tested Laplacians match reconstruction exactly, while elasticity differences remain at rounding scale.

For fixed coefficients, maintaining the combined time-step matrix avoids a global combination pass.
In the four-step refinement replay, Direct Incremental with bulk initialization reduces the measured total of setup, preparation, preconditioning, and solving by 2.7\% compared with compact rebuild.
This result is specific to the tested replay and includes the cost of applying the reserved matrix.

These results show why assembly time alone is not enough to choose an update policy.
Setup, stored zeros, and solver work can limit the benefit.
Future work could study compiled implementations, warm starts and preconditioner reuse, connections to system-level updates such as AMPS \cite{yeung2018amps}, and element updates for nonlinear materials.

\begin{acks}
D.H.\ was supported by NSF grant number 2450401.
\end{acks}

\bibliographystyle{ACM-Reference-Format}
\bibliography{references}

\end{document}